\documentclass[conference]{IEEEtran}
\usepackage{cite}
\usepackage{amsmath,amssymb,amsfonts}
\usepackage{algorithmic}
\usepackage{graphicx}
\usepackage{textcomp}
\usepackage{xcolor}
\usepackage{hyperref}
\usepackage{stfloats}
\usepackage{xfrac}
\usepackage{listings}
\usepackage{enumerate}
\usepackage{enumitem}
\usepackage{caption}
\usepackage[scientific-notation=true,round-precision=3,round-mode=figures]{siunitx}
\usepackage{color,soul}
\usepackage{url}
\usepackage{multirow}
\usepackage[multiple]{footmisc}
\usepackage{cleveref}

\def\BibTeX{{\rm B\kern-.05em{\sc i\kern-.025em b}\kern-.08em
    T\kern-.1667em\lower.7ex\hbox{E}\kern-.125emX}}
\usepackage{comment}

\newcommand{\ie}{\emph{i.e.,\,}}

\makeatletter

\newcommand\Autoref[1]{\@first@ref#1,@}
\def\@throw@dot#1.#2@{#1}%
\def\@set@refname#1{%
    \edef\@tmp{\getrefbykeydefault{#1}{anchor}{}}%
    \xdef\@tmp{\expandafter\@throw@dot\@tmp.@}%
    \ltx@IfUndefined{\@tmp autorefnameplural}%
         {\def\@refname{\@nameuse{\@tmp autorefname}s}}%
         {\def\@refname{\@nameuse{\@tmp autorefnameplural}}}%
}
\def\@first@ref#1,#2{%
  \ifx#2@\autoref{#1}\let\@nextref\@gobble%
  \else%
    \@set@refname{#1}%
    \@refname~\ref{#1}%
    \let\@nextref\@next@ref%
  \fi%
  \@nextref#2%
}
\def\@next@ref#1,#2{%
   \ifx#2@ and~\ref{#1}\let\@nextref\@gobble%
   \else, \ref{#1}%
   \fi%
   \@nextref#2%
}

\makeatother

\begin{document}

\title{Characterizing Orphan Transactions\\ in the Bitcoin Network}

\author{
\IEEEauthorblockN{Muhammad Anas Imtiaz}
\IEEEauthorblockA{\textit{ECE Department} \\
\textit{Boston University}\\
Boston, MA 02215, USA \\
\href{mailto:maimtiaz@bu.edu}{maimtiaz@bu.edu}}
\and
\IEEEauthorblockN{David Starobinski}
\IEEEauthorblockA{\textit{ECE Department} \\
\textit{Boston University}\\
Boston, MA 02215, USA \\
\href{mailto:staro@bu.edu}{staro@bu.edu}}
\and
\IEEEauthorblockN{Ari Trachtenberg}
\IEEEauthorblockA{\textit{ECE Department} \\
\textit{Boston University}\\
Boston, MA 02215, USA \\
\href{mailto:trachen@bu.edu}{trachen@bu.edu}}
}

\IEEEoverridecommandlockouts
\IEEEpubid{\makebox[\columnwidth]{978-1-7281-6680-3/20/\$31.00 \copyright 2020 IEEE \hfill} \hspace{\columnsep}\makebox[\columnwidth]{ }}

\maketitle

\IEEEpubidadjcol

\begin{abstract}
    Orphan transactions are those whose parental income sources are missing at the time that they are processed.
    These transactions typically languish in a local buffer until evicted or all their parents are
    discovered, at which point they may be propagated further.  So far, there has been little work in the literature on characterizing the nature and impact of such orphans. Yet, it is intuitive that they
    should affect performance of the Bitcoin network. %
    This work thus seeks to methodically research such effects through a measurement campaign of orphan transactions
    on live Bitcoin nodes. 
    Our data show that, surprisingly, orphan transactions tend to have fewer parents on average than non-orphan transactions. 
    The salient features of their missing parents are a lower fee, a larger size, and a lower transaction fee per byte than all other received transactions.
    We also find out that the network overhead incurred by these orphan transactions can be significant, exceeding 17\% when
    using the default orphan memory pool size (\ie 100 transactions).  However, this overhead can be made negligible, without
    significant computational or memory demands, if the pool size is simply increased to 1000 transactions.
\end{abstract}


\section{Introduction}
With a market cap of over $135$ billion US dollars~\cite{btc_market_cap}, the Bitcoin cryptocurrency has come a long
way since its introduction as a peer-to-peer, electronic cash system by Satoshi Nakamoto in $2008$~\cite{nakamoto2008bitcoin}. 
Nodes within the Bitcoin network exchange \emph{transactions} to record purchases and sales using Bitcoin currency, one unit of
which is further subdivided into 100 million \emph{satoshis}.  After such a transaction is created,
it is propagated through the Bitcoin network, whose nodes add it to their local memory buffer called a \emph{mempool}.  Transactions
stay in the mempool until confirmed by a Bitcoin miner~\cite{unconfirmed_tx} and added to a block in the common ledger known as a \emph{blockchain}. Every day, hundreds of thousands of transactions are created and confirmed in the Bitcoin network~\cite{daily_confiremd_txs}, resulting in a total of over $480$ million transactions since its inception~\cite{total_txs}.

    Before relaying a transaction to its peers, a node in the Bitcoin network must confirm that the transaction has verified currency input from its \emph{parent} transactions. %
    If a transaction's parents are not in the node's mempool or local blockchain, then the 
    transaction is classified an \emph{orphan}, and it is not relayed further until the parents arrive.  We seek to more precisely understand the context under which a transaction becomes an orphan, including the properties of parent transactions that produce this effect.

\subsection{History}
    Bitcoin transactions have received a fair amount of attention in the literature. Subset of this work have focused on elements such as an analysis of the transaction graphs~\cite{ron2013quantitative, ober2013structure, fleder2015bitcoin, di2015bitconeview, moser2013inquiry, greaves2015using, mcginn2016visualizing}, security of transactions~\cite{androulaki2013evaluating, ruffing2017valueshuffle, herrera2016privacy, wang2017preserving, liu2018unlinkable, meiklejohn2018mobius}, studies on transaction confirmation times~\cite{kawase2017transaction, sompolinsky2013accelerating, kasahara2019effect, zhu2016interactive}, and the like. %

    Understanding the properties and behavior of orphan transactions, however, is a largely unexplored field.  The closest works have
    been on utilizing orphan transactions as a side-channel for topology inference~\cite{delgado2019txprobe}, and for denial of service attacks on the Bitcoin network~\cite{miller2015shadow, cve_wiki}. However, many of the performance questions regarding orphan transactions remain: To \emph{what} extent orphan transactions are prevalent in the Bitcoin network? What are the factors that make a transaction orphan? What is the impact of an orphan transaction on the performance of the Bitcoin ecosystem? Does an orphan transaction incur latency or communication overhead? If so, can one reduce this overhead? There exists no work, to the best of our knowledge, that reasonably answers these questions.

    \subsection{Contributions}
    Our first contribution in this paper is to characterize orphan transactions in the Bitcoin network and identify the environment that
    produces them, based on a data set of $\num{4200015}$ \emph{unique} transactions ($\num{87125}$ of which are orphans) received over the measurement period. We discover that the intuition that orphan transactions may have larger numbers of parents than non-orphans (presumably resulting in a greater probability that one of the parents is missing) is misleading.
    Indeed, orphan transactions generally have \emph{fewer} parents than all other transactions received during our measurements, averaging $1.18$ parents (orphans) versus $2.20$ (non-orphans). %
    We conclude that the number of parents does not suitably distinguish between orphan and non-orphan transactions.

    We then consider other metrics (\ie transaction fee, transaction size, and transaction fee per byte) to discern the distinction
    between these two types of transactions.  
    Our analysis shows that missing parents of orphan transactions have smaller fees and larger size than all other received transactions. More precisely, a missing parent of an orphan transaction has an average transaction fee of $\num{5560.80}$ satoshis, %
    and an average transaction size of $\num{528.70}$ bytes. By comparison, all other transactions have an average transaction fee of $\num{9911.61}$ satoshis and transaction size of $\num{480.31}$ bytes. Finally, we find that, on an individual level, missing parents of orphan transactions pay a fee of $6.25$ satoshis per byte versus $21.73$ satoshis per byte for all received transactions. 
    As a result, transactions with a smaller fee per byte are more likely to go missing and render their descendent transactions orphans.

    Our second contribution is to study the impact of network and performance overhead caused by orphan transactions. We thus collect data from live nodes in the Bitcoin network with various orphan pool sizes (including the default of $100$). Our measurements show that orphan transactions incur a significant network overhead (\ie number of bytes received by their node) when the orphan pool size is smaller. In effect, the pool fills up and transactions in the orphan pool are rapidly evicted to make room for new orphan transactions. As such, an orphan transaction may be added to the orphan pool multiple times as it is announced by different peers. We show that by slightly increasing the orphan pool size to $1000$ transactions, we can dramatically reduce this network overhead without a distinguishable effect on node performance (in terms of computation and memory). 

\subsection{Road map}
    The rest of this paper is organized as follows: In Section~\ref{sec:background}, we present preliminary background and related work. In Section~\ref{sec:charac}, we characterize orphan transactions by studying the properties of their parents. We show the impact of orphan transactions with varying orphan pool sizes in Section~\ref{sec:comparison}. Section~\ref{sec:conclusion} concludes the paper and discusses potential areas for future work. 

\section{Background and related work}\label{sec:background}

    In this section, we provide relevant background material on the orphan transactions followed by a
    discussion
    of related work.

    \subsection{Orphan transactions}\label{subsec:orphantxs}

        A Bitcoin node may receive a transaction that spends income from one or more yet unseen parent transactions (\ie the parents are neither included in any of the previous blocks of the Bitcoin blockchain nor exist in the node's mempool). The node cannot accept the newly received transaction into its mempool until it can verify that the transaction spends valid Bitcoin, and it thus requests the missing parents from the peer that originally sent the transaction.  In the meanwhile, the transaction is classified as an \emph{orphan} transaction and added to an \emph{orphan pool} that is maintained in the \texttt{mapOrphanTransactions} data structure
        in the Bitcoin core software. 

        Once the orphan transaction is added to the orphan pool, there are six cases that can cause its removal (corresponding to lines $76$, $2331$, $2326{-}2330$, $1609{-}1620$, $876{-}906$, $800{-}806$, $40$, $784{-}794$, $627$, $757{-}771$, $1624{-}1632$, and $1608$ in the core implementation of \texttt{netprocessing.cpp}~\cite{BitcoinNetProcessing}):
        
        \begin{enumerate}[label=\arabic*.]
            \item \textbf{Parent transactions received.} The node receives a parent it requested from its peer. It then processes any orphan transactions that depend on the newly received transaction. All transactions that are no longer orphan are removed from the orphan pool and added to the mempool.
            \item \textbf{Parent transactions in block.} The node receives a new block which contains the missing parents needed to verify the orphan transaction. The node iterates over the transactions in the block and removes any orphan transactions from the orphan pool that depend on the former and add it to the mempool.
            \item \textbf{Orphan pool full.} By default, the size of the orphan pool is capped to a maximum of $100$ orphan transactions. 
            When the orphan pool is full, an orphan transaction is chosen at random and removed from the pool, and
            this transaction is not added to the mempool. The maximum size of the orphan pool can be modified at startup by using the \texttt{-maxorphantx} argument when running \texttt{bitcoind} or \texttt{bitcoin-qt}, or set in the \texttt{bitcoin.conf} configuration file~\cite{bitcoinconf}.
            \item \textbf{Timeout.} By default, an orphan transaction \emph{expires} and is removed after $20$ contiguous minutes  in the orphan pool.
            \item \textbf{Invalid orphan transaction.} The node deems that an orphan transaction is invalid when the missing parents of the orphan transaction have been received, but the orphan transaction itself may be non-standard or not have sufficient fee. Thus, this orphan transaction is not accepted to the mempool. Furthermore, not only the orphan transaction is removed from the orphan pool, but also the peer that originally sent the orphan transaction is punished, \ie no further transactions are accepted to the mempool from the peer in the current round.
            \item \textbf{Peer disconnected.} When a peer disconnects from a node, all orphan transactions sent by this peer are removed from the orphan pool in the finalization step.
            This is likely because the node no longer expects to receive the parents it requested from the peer. The orphan transaction is not added to the mempool.
        \end{enumerate}

        A transaction may get \emph{stuck}~\cite{stuck_tx} in mempools of nodes due to low transaction fees. That is, the transaction is not included in blocks and faces delays in confirmation. 
        Bitcoin does allow the transactions to be modified to increase the fee~\cite{bip125}, and the originator of the transaction may add a new input, \ie a new parent, as a spending source for the increased fee. The transaction may become orphaned if the new input is missing from the receiving node's mempool or local blockchain, and this transaction is then added to the orphan pool. We do not classify such orphan transactions separately because they do make it to the orphan pool.

    \subsection{Related work}

        To the best of our knowledge, there is very little work in the Bitcoin research literature regarding orphan transactions. Nevertheless, the few works that do consider them highlight the potential
        value
        of the area, and the need for further work.

        Miller and Jansen~\cite{miller2015shadow} take advantage of the fact that in the older version of Bitcoin (\ie v0.9.2), the protocol did not keep track of the peer that sent an orphan transaction. They propose that an adversary can leverage this vulnerability to mount a denial of service attack by sending a large number of orphan transactions to the victim node. The latter would be stuck verifying the transaction signatures of orphan transactions for a long time. However, this threat model is outdated since, in the current version, the Bitcoin protocol does keep track of the sender of an orphan transaction. The work also does not present a characterization of the orphan transactions.

        Delgado-Segura et. al.~\cite{delgado2019txprobe} present \emph{TxProbe}, a technique that makes use of orphan transactions to deduce the topology of the Bitcoin network. In this approach, an adversary creates a pair of double-spending transactions, and propagates each to a different node. The nodes try to propagate the double-spending orphan transaction to one another, if there exists an edge between the two. However, each of the receiving node rejects the incoming transaction as an invalid double-spending transaction. The adversary then sends a transaction that spends from one of the double-spending transactions to the node that received the corresponding double-spending transaction. This latter node will propagate the new transaction to the second node, if there exists an edge between the two. However, the second node will add the new transaction to its orphan pool, since it already rejected its parent earlier. The adversary can then probe the second node for the orphan transaction to establish a side-channel:  if the node responds with the orphan transaction, the adversary deduces that there exists an edge between the two nodes that received the pair of double-spending transactions. The authors then extend this basic approach to a larger Bitcoin graph. Though this work presents an interesting side-channel in the Bitcoin network, it also does not characterize orphan transactions.

        Earlier version of Bitcoin software did not place a limit on the number of orphan transactions that a node can store. Thus, an adversary could launch a denial of service attack by sending a large number of orphan transactions to a victim node, causing memory exhaustion and system failure. Furthermore, the Bitcoin software did not contain validation checks for the size of an orphan transaction. Hence, an adversary could create an orphan transaction with an arbitrarily large size and cause memory exhaustion at the victim node~\cite{cve_blog}. Both of these vulnerabilities were responsibly reported 
        and fixed~{\cite{dos_fix, cve_wiki}}. While our work proposes increasing the size of the orphan pool, the current validation checks should ensure that this change will not enable denial of service attacks.

\section{Characterization of orphan transactions}\label{sec:charac}

    We next detail our approaches toward characterizing the orphan transactions in the Bitcoin network.  We begin with a presentation of our set up for data collection. Since a transaction becomes orphan due to the absence of one or more parents, we next focus on determining the characteristics of these missing parents. In particular, we compare the number of parents of orphan transactions with number of parents of all non-orphan transactions. Thereafter, we consider the differences between the transaction fee, transaction size, and transaction fee per byte of the missing parents of orphan transactions versus all other transactions.

    \subsection{Measurement setup}\label{subsec:charac_meas_setup}

        We run two live full nodes $N_1$ and $N_2$ as part of the Bitcoin network, with the aim of collecting data for characterizing orphan transactions. Both nodes execute Bitcoin Core v$0.18$~\cite{bitcoin_modified} on the Linux Ubuntu $18.04.2$ LTS distribution, running on Dell Inspiron $3670$ desktops, each equipped with an $8^{\text{th}}$ Generation Intel\textregistered~Core i$5{-}8400$ processor ($9$~MB cache, up to $4.0$~GHz), $1$~TB HDD and $12$~GB RAM.  The nodes are connected to the Bitcoin network at all times with the default orphan pool size of $100$. We collect relevant data, such as arrival of transactions, addition of transactions to the orphan pool, and the like, with the help of a log-to-file system~\cite{imtiaz2019churn}, for roughly $2$ weeks over two rounds (November $18$, $2019$ $11{:}00$ AM to November $25$, $2019$ $10{:}59$ AM, and November $25$, $2019$ $11{:}00$ AM to December $02$, $2019$ $10{:}59$ AM).

    \subsection{Number of parents}

        \begin{figure}[t!]
            \centering
            \includegraphics[width=\columnwidth, clip,trim=0.5cm 0cm 5cm 4cm]{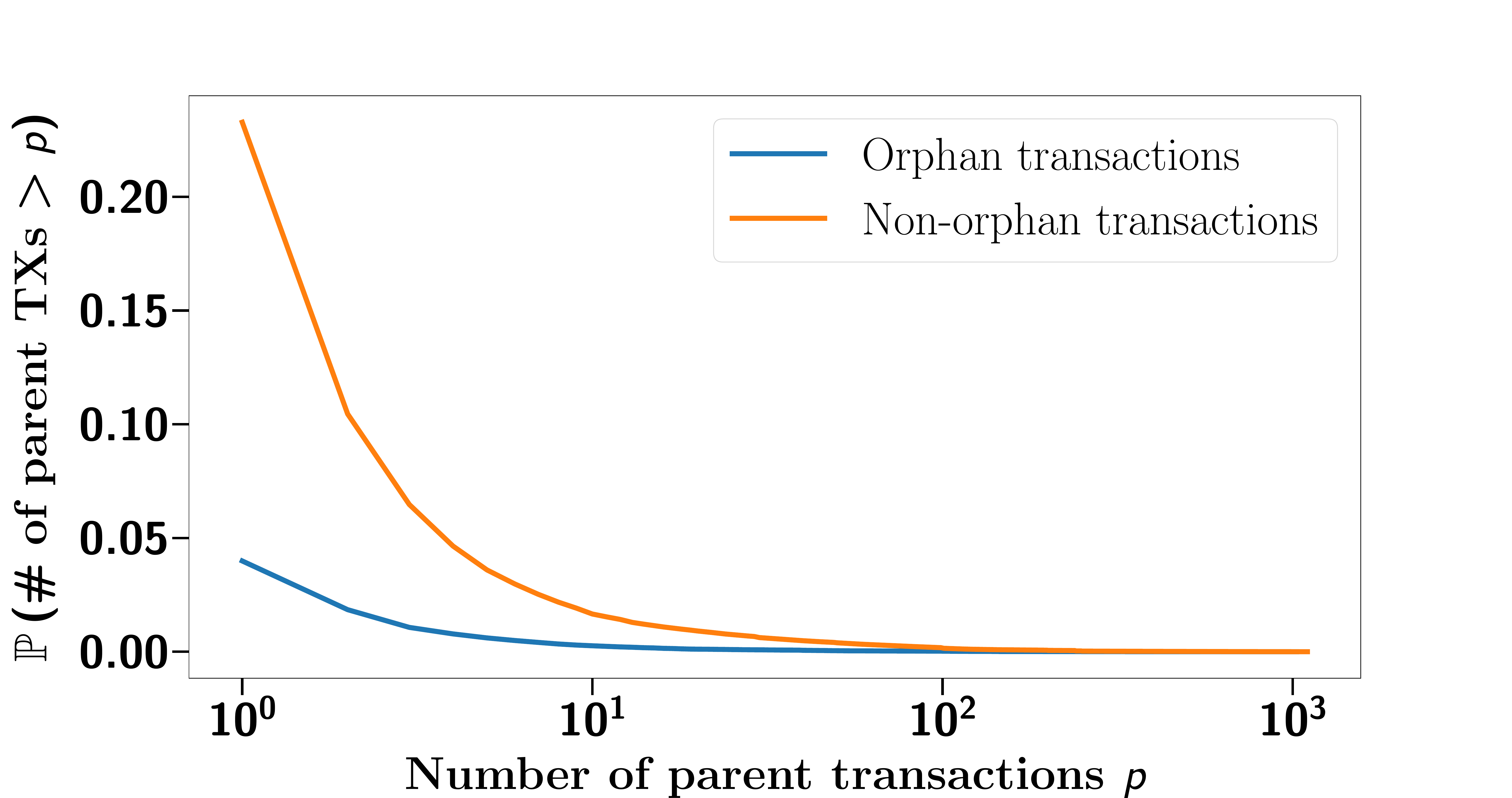}
            \caption{Empirical complementary cumulative distribution function (CCDF) of (i)~the number of parents of orphan transactions and (ii)~number of parents of non-orphan transactions. In general, orphan transactions have fewer parents.}
            \label{fig:num_parents}
        \end{figure}

        Our first conjecture is that a transaction with a large number of parents may be more likely to miss one or more parents than a transaction with, say, only a couple of parents. To this effect, we compare the number of parents of orphan transaction with the number of parents of all other
        non-orphan transactions.

        During the measurement period, the nodes receive an aggregate of $\num{4200015}$ \emph{unique} transactions with $\num{9232112}$ parents. Of these, $\num{87125}$ are orphan transactions with $\num{103057}$ parents. These orphan transactions have an aggregate of $\num{87144}$ parents missing across the nodes. These nodes miss, on average, $1.23$ parents per orphan transaction with a standard deviation of $4.68$ parents. While only just above $2\%$ of the received transactions become orphan, the total number is still significant. We observe that, on average, roughly $56\%$ of the orphan transactions make it into the blocks received during the measurement period.
        
        \figurename~\ref{fig:num_parents} shows the complementary cumulative distribution functions (CCDF) of the number of parents of orphan transactions, and the CCDF of the number of parents of non-orphan transactions. We observe that our conjecture is flipped - the orphan transactions have a \emph{smaller} number of parents. Indeed, only about $4\%$ of orphan transactions have more than one parents, whereas roughly $25\%$ of non-orphan transactions have more than one parent.
        
        The most parents of an orphan transaction is $\num{1027}$, whereas this number is $\num{1102}$ for non-orphan transactions.
        On average, an orphan transaction has $1.18$ parents with a standard deviation of $4.78$ transactions. On the other hand, a non-orphan transaction has, on average, $2.20$ parents with a standard deviation of $11.84$ transactions.

        Surprisingly, orphan transactions do not necessarily have more parents than non-orphan transactions, and we are left to rely on other statistics, presented in the next few sections, to characterize the orphan transactions.

    \subsection{Transaction fee of missing parents}\label{subsec:tx_fee}

        \begin{figure}[t!]
            \centering
            \includegraphics[width=\columnwidth, clip,trim=1.65cm 0cm 5cm 3.65cm]{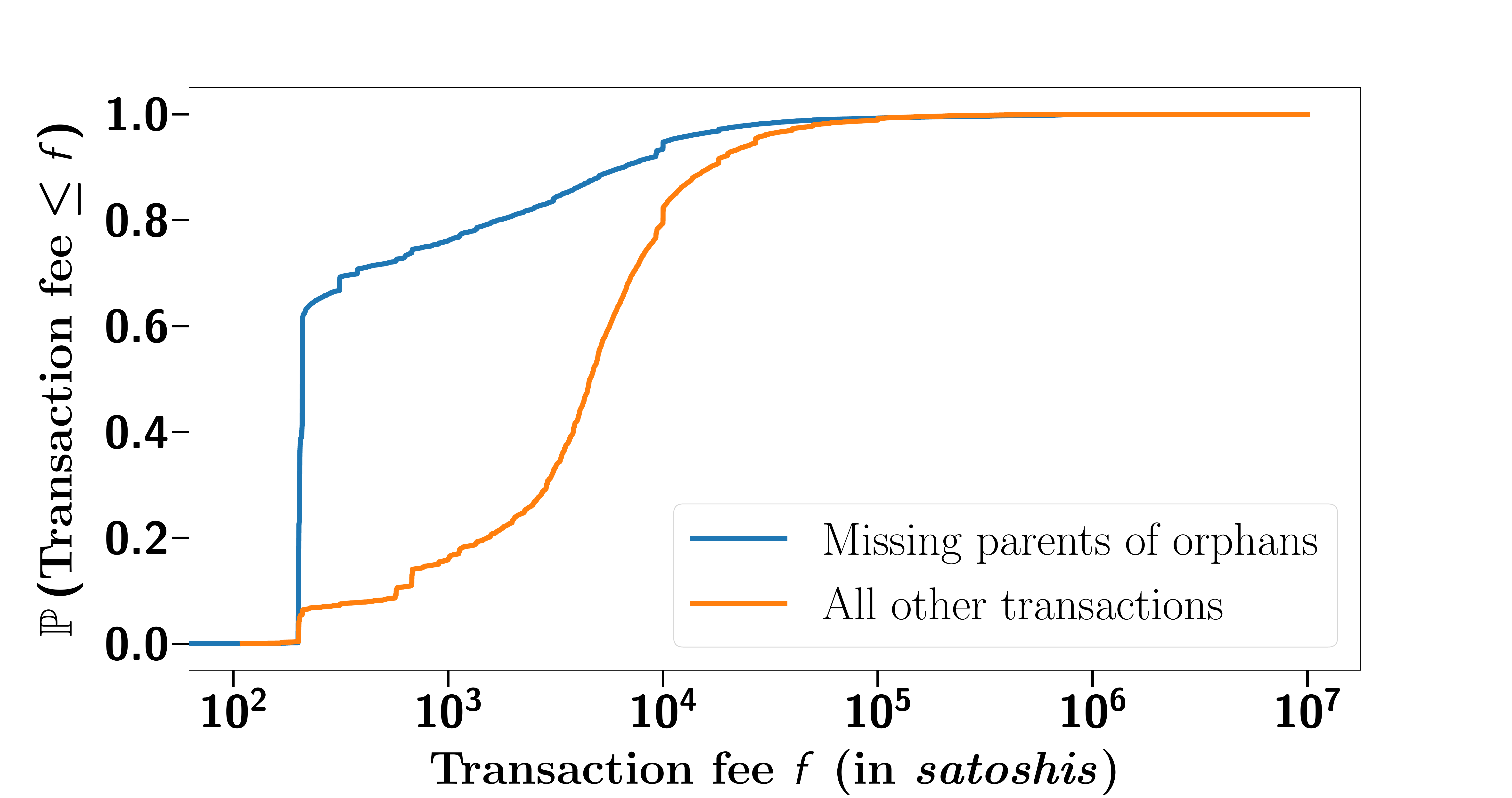}
            \caption{Cumulative distribution functions (CDFs) of transaction fee of missing parents of orphan transactions, and transaction fee of all other transactions.}
            \label{fig:tx_fees}
        \end{figure}

        For each incoming transaction that is orphaned, we log the missing parent(s) that results in the transaction becoming orphan. We analyze and compare the transaction fees of these missing parents with all
        other transactions received by our nodes that are not a missing parent of an orphan transaction. We query the database maintained by the Bitcoin software for relevant data on transactions. Out of $8.71\times 10^4$ missing parents, only about 
        $3\%$ are still missing by the end of the measurement period. Henceforth, we assume that this relatively small
        fraction does not pose a bias towards our findings.

        \figurename~\ref{fig:tx_fees} shows the cumulative distribution functions (CDFs) of transaction fees (in \emph{satoshis}) of missing parents, and the CDF of transaction fees (in \emph{satoshis}) of all other transactions received by the nodes. The figure shows that a majority of the missing parents have a lower transaction fee compared to all other transactions received. Indeed, $50\%$ of missing parents have a transaction fee smaller than $210$ satoshis. On the other hand, fewer than $6\%$ of all other transactions have a transaction fee of smaller than $210$ satoshis.

        In fact, the average transaction fee of a missing parent is $\num{5560.80}$ satoshis with a standard deviation of $\num{71725.30}$ satoshis. In comparison, the average transaction fee of all other transactions is $\num{9911.61}$ satoshis with a standard deviation of $\num{55262.91}$ satoshis. Interestingly, $18$ of the missing parents have \emph{no} transaction fee  at all (\ie $0$ satoshis), whereas all other transactions received have a non-zero transaction fee.

        Therefore, a transaction is likely to become an orphan, if its missing parent has a transaction fee lower than that of other transactions. As a future work, it would be interesting to deduce if there exists a threshold for the transaction fee below which all transactions become missing, \ie they are not relayed by the network.

    \subsection{Transaction size of missing parents}\label{subsec:tx_size}

        \begin{figure}[t!]
            \centering
            \includegraphics[width=\columnwidth, clip,trim=1.65cm 0cm 5cm 3.50cm]{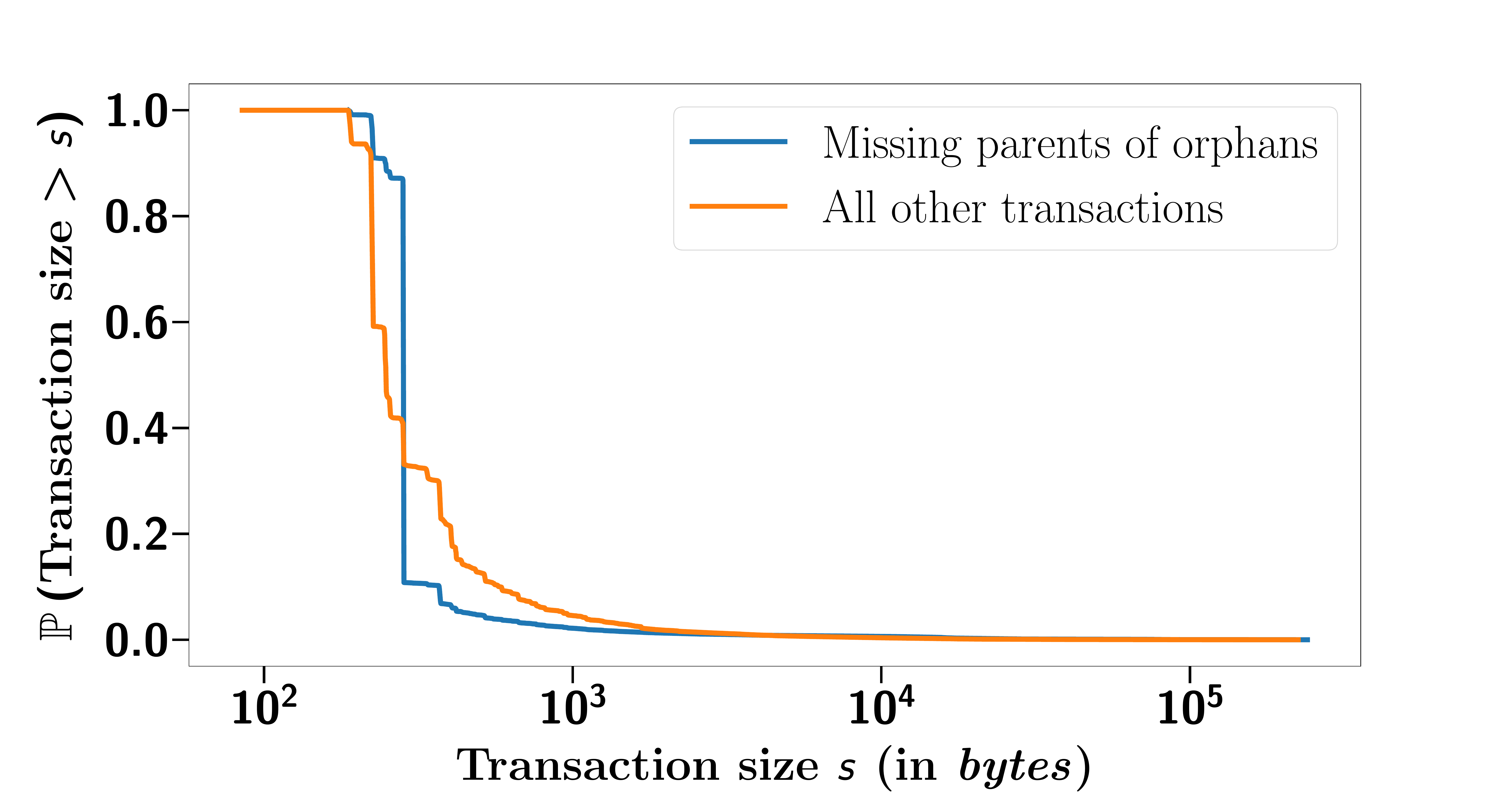}
            \caption{CCDFs of transaction size of missing parents of orphan transactions, and transaction size of all other transactions.}
            \label{fig:tx_size}
        \end{figure}

        We next compare the sizes of missing parents of orphan transactions with the sizes of all other transactions. Do the missing parents of orphan transactions have a larger size than an average transaction?

        \figurename~\ref{fig:tx_size} shows the CCDF of the size of missing parents of orphan transactions (in bytes) and the CCDF of the size of all other transactions. The figure shows that missing parents usually have a larger size than all other transactions. Roughly $90\%$ of the missing parents have a size larger than $250$ bytes, whereas only about $45\%$ of all other transactions have a size larger than $250$ bytes.

        Missing parents of orphan transactions have a size between $\num{188}$ and $\num{240208}$ bytes. By comparison, all other transactions have a size in the range of $\num{85}$ to $\num{224183}$ bytes. In fact, on average, missing parents have a size of $\num{528.70}$ bytes with a standard deviation of $\num{4018.97}$ bytes. On the other hand, all other transactions have, on average, a size of $\num{480.31}$ bytes with a standard deviation of $\num{2120.40}$ bytes.

        The statistics in this section show that the missing parents of orphan transaction have, on average, a larger transaction size than all other transactions. As in the previous section, we leave to future work the question whether there exists a size threshold above which  transactions stop being propagated through the network.

    \subsection{Relating transaction fee to size of missing parents}

        We showed in \Autoref{subsec:tx_fee, subsec:tx_size}
        that, in aggregate, missing parents tend to have a lower fee and a larger size than the average received transactions. However, it would be interesting to see if there exists a relation between the fee and size of each individual transaction.

\begin{figure}[t!]
            \centering
            \includegraphics[width=\columnwidth, clip,trim=1.65cm 0cm 5cm 3.65cm]{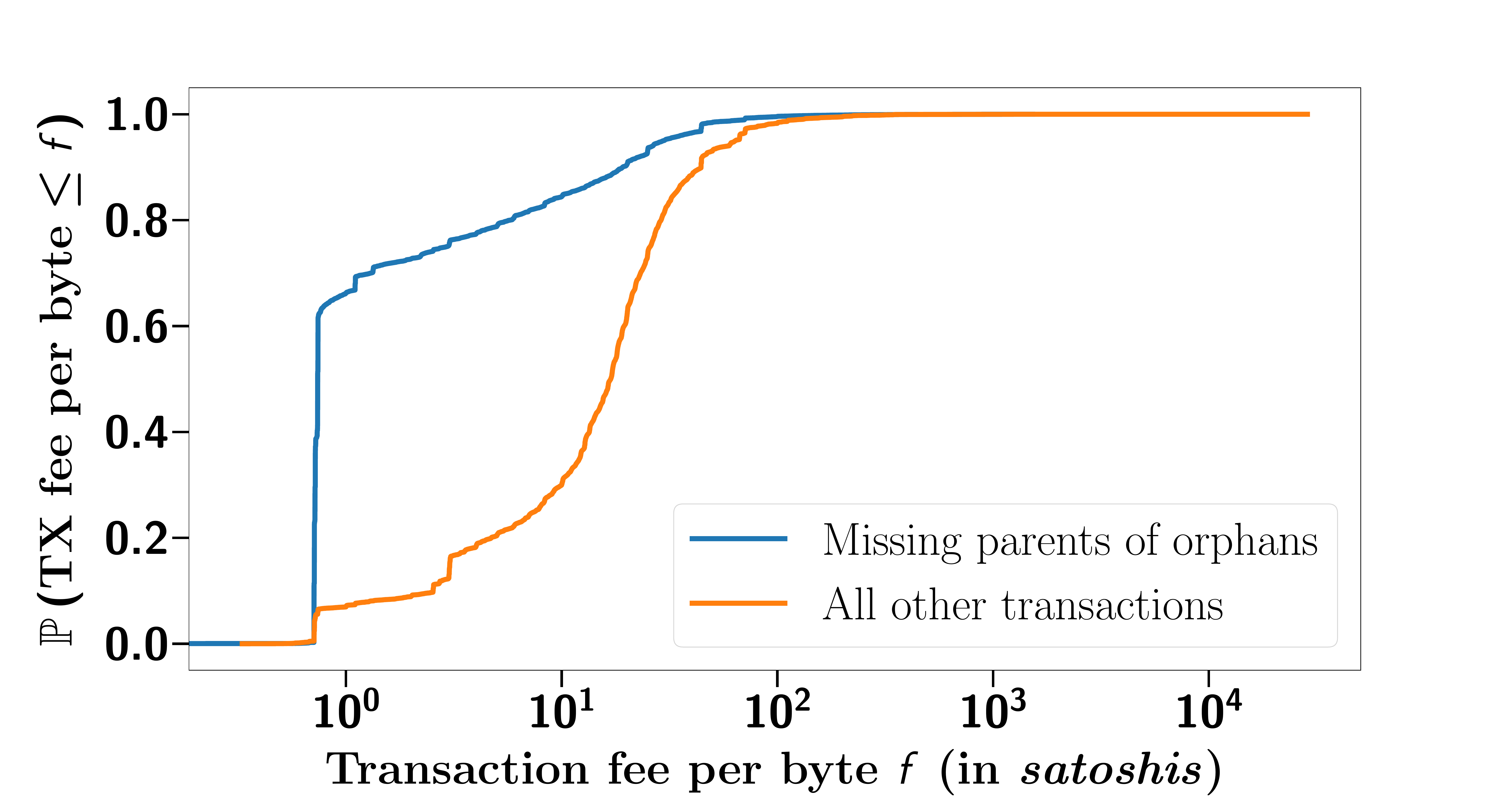}
            \caption{CDFs of transaction fee per byte of missing parents of orphan transactions, and transaction fee per byte of all other transactions.}
            \label{fig:tx_fee_per_byte}
        \end{figure}

        To this end, \figurename~\ref{fig:tx_fee_per_byte} shows the CDF of transaction fee \emph{per byte} (in satoshis) of missing parents and the CDF of transaction fee per byte of all transactions received. The figure shows that the missing parents generally have a lower transaction fee per byte when compared to all received transactions. Indeed, $80\%$ of missing parents have a transaction fee per byte of $5.97$ satoshis or less, whereas roughly $78\%$ of all received transactions have a transaction fee per byte higher than $5.97$ satoshis.
        On average, missing parents have a transaction fee per byte of $6.25$ satoshis with a standard deviation of $21.52$ satoshis. On the other hand, all received transactions have a transaction fee per byte of $21.73$ satoshis with a standard deviation of $47.13$ satoshis. %

        Our data thus show that individual missing parents have a low transaction fee per byte. This could be because transactions with lower fees may not get properly propagated through the Bitcoin network~\cite{tx_low_fee_prop}, possibly because of configurable mempool size~\cite{mempool_size}. Note that nodes may choose not to accept transactions with a low transaction fee per byte to their mempool, and thereby not propagate them further~\cite{jiang2019bitcoin}.

\section{Comparison of orphan transaction behavior with different orphan pool sizes}\label{sec:comparison}

    We next characterize the network and performance overhead incurred by orphan transactions, looking at both the default orphan pool size of $100$ transactions, and various alternative pool sizes.  We begin with a presentation of our extended measurement setup, followed by an investigation of the network overhead under additions and removals of orphan transactions fir different orphan pool sizes. Finally, we discuss performance overhead that a larger orphan pool size may present.

    \subsection{Measurement setup}

        \begin{figure}[t!]
            \centering
            \includegraphics[width=\columnwidth, clip,trim=18.5cm 1.5cm 9cm 3.5cm]{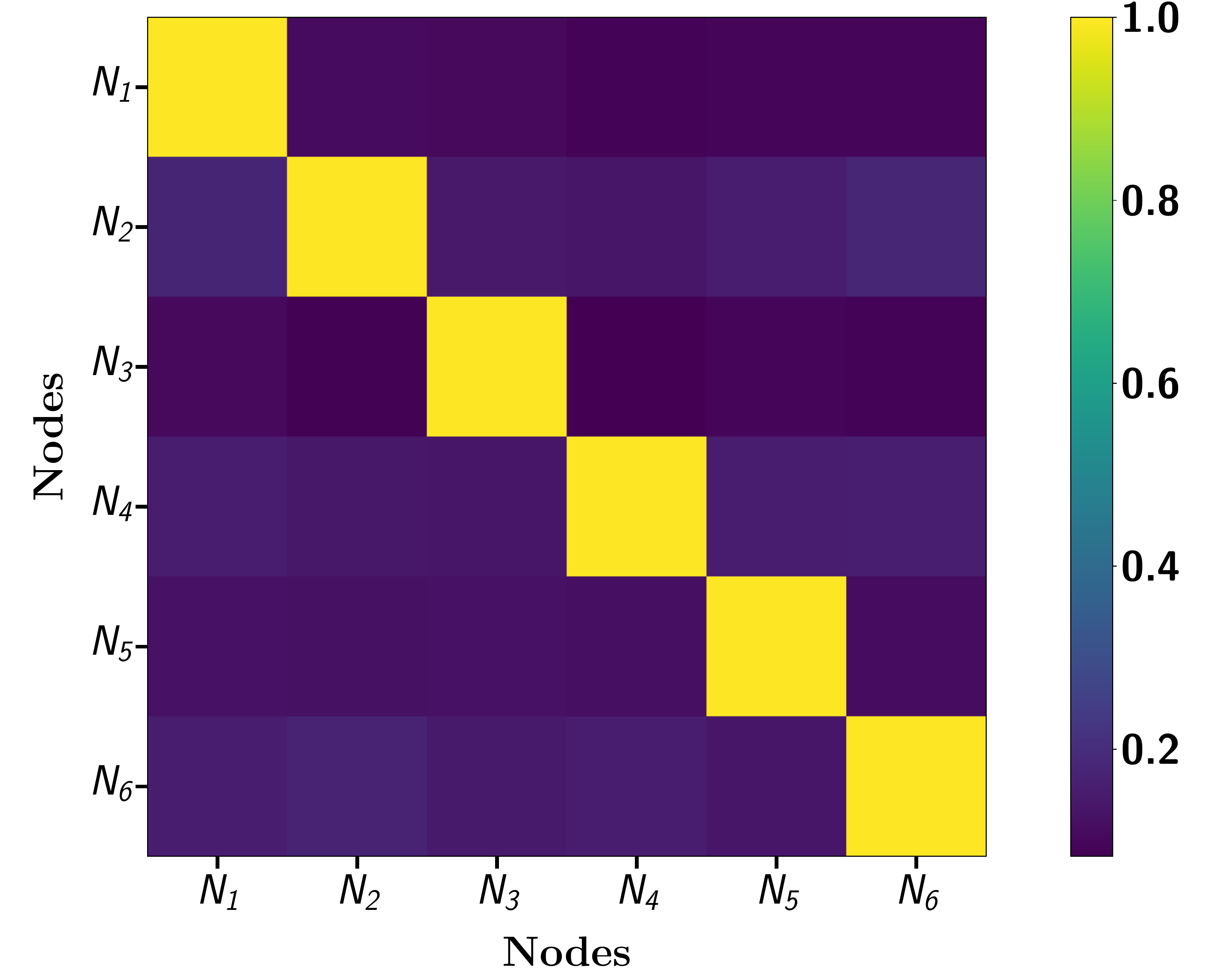}
            \caption{Similarity matrix depicting average number of common peers across nodes during the first round of measurement period.}
            \label{fig:sim_matrix_r1}
        \end{figure}

        \begin{figure}[t!]
            \centering
            \includegraphics[width=\columnwidth, clip,trim=18.5cm 1.5cm 9cm 3.5cm]{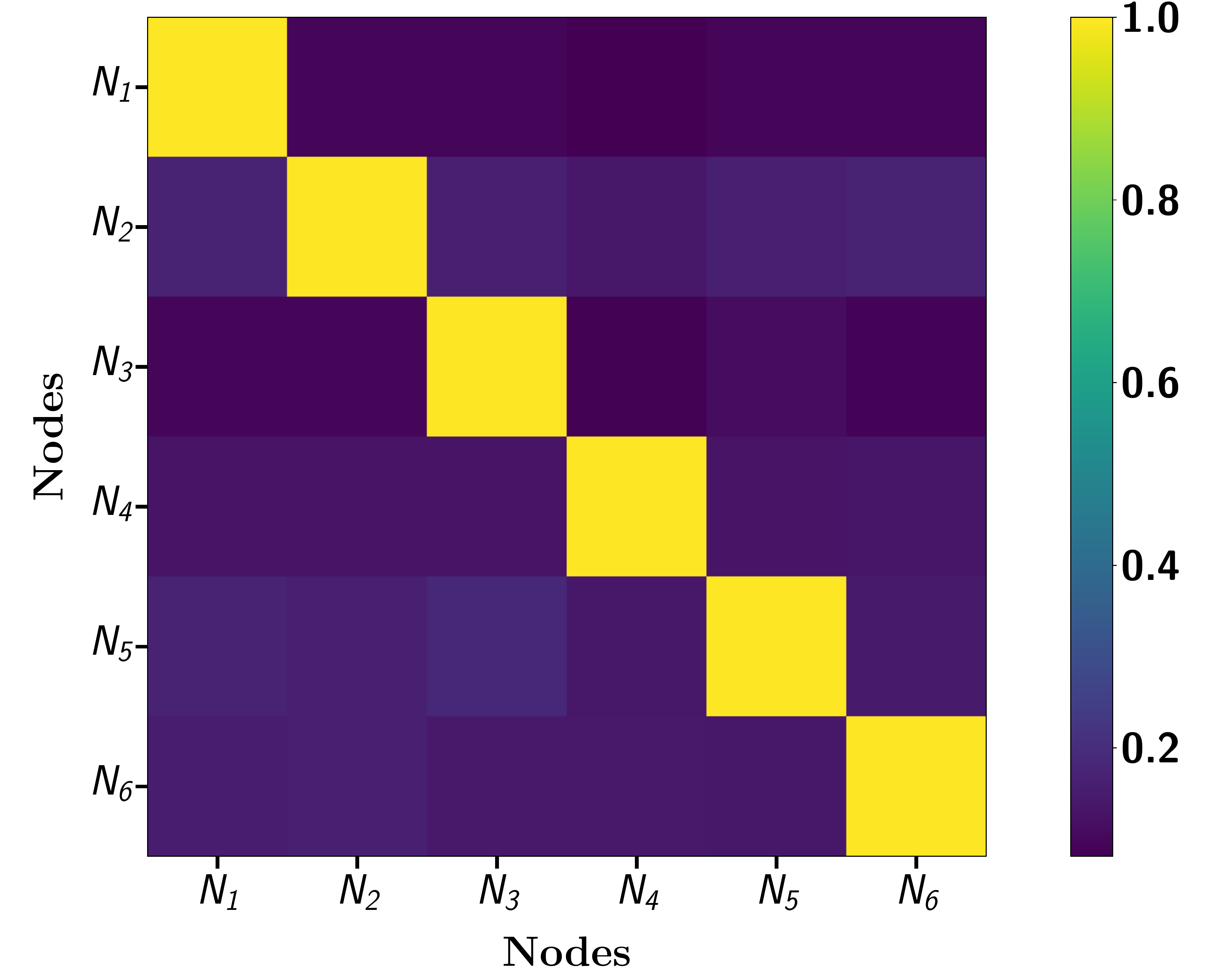}
            \caption{Similarity matrix depicting average number of common peers across nodes during the second round of measurement period.}
            \label{fig:sim_matrix_r2}
        \end{figure}

        We extend our measurement setup from \autoref{subsec:charac_meas_setup} to six live full nodes, running with identical hardware and software specifications as before. We run two rounds of experiments. In the first round, which runs from November $18$, $2019$ $11{:}00$ AM to November $25$, $2019$ $10{:}59$ AM, two nodes are configured with a default orphan pool size of $100$ transactions (nodes $N_1$ and $N_2$), two nodes with an orphan pool size of $20$ transactions (nodes $N_3$ and $N_4$), and the remaining two nodes with an orphan pool size of $50$ transactions (nodes $N_5$ and $N_6$). In the second round, which runs from November $25$, $2019$ $11{:}00$ AM to December $02$, $2019$ $10{:}59$ AM, two nodes are configured with a default orphan pool size of $100$ transactions (nodes $N_1$ and $N_2$), two nodes with an orphan pool size of $500$ transactions (nodes $N_3$ and $N_4$), and the remaining two nodes with an orphan pool size of $1000$ transactions (nodes $N_5$ and $N_6$). We have made all relevant logs generated during the experiments open source and accessible on GitHub~{\cite{bitcoin-logs}}.

        Since our nodes are co-located, we want to verify that the nodes connect independently to outside peers in the network, and that our co-location does not impose a bias in the measurements. We achieve this by recording a node's connected peers over time, in one second intervals. We then check for common peers amongst the nodes throughout the measurement period, \ie both the first and the second rounds.

        \figurename~\ref{fig:sim_matrix_r1} and \figurename~\ref{fig:sim_matrix_r2} show the common peers amongst nodes during the measurement period (\ie the first and second rounds of measurement respectively) as similarity matrices. A similarity score of $1.0$ between two nodes indicates that both nodes have exactly the same peers; a similarity score of $0.0$ indicates that the corresponding nodes have no common peers. The matrices in the figures qualitatively suggest that the six nodes have a very low number of peers in common, and therefore, do not present bias towards measurements.

        In fact, the maximum number of peers that all six nodes have in common during the first round of measurements is $11$ peers out of a maximum of $124$ peers. On average, at any second during the measurement period, all six nodes have $8.30$ peers in common with a standard deviation of $1.04$ peers. Similarly, during the second round of measurements, the maximum number of peers that all six nodes have in common is $11$ peers out of a maximum of $124$ peers. On average, at any second during the measurement period, all six nodes have $8.51$ peers in common with a standard deviation of $0.92$ peers. These statistics confirm that nodes largely connect to, and interact with peers independently.

        \begin{figure*}[t!]
            \centering
            \includegraphics[width=\linewidth, clip, trim=0.25cm 2.75cm 0.25cm 4cm]{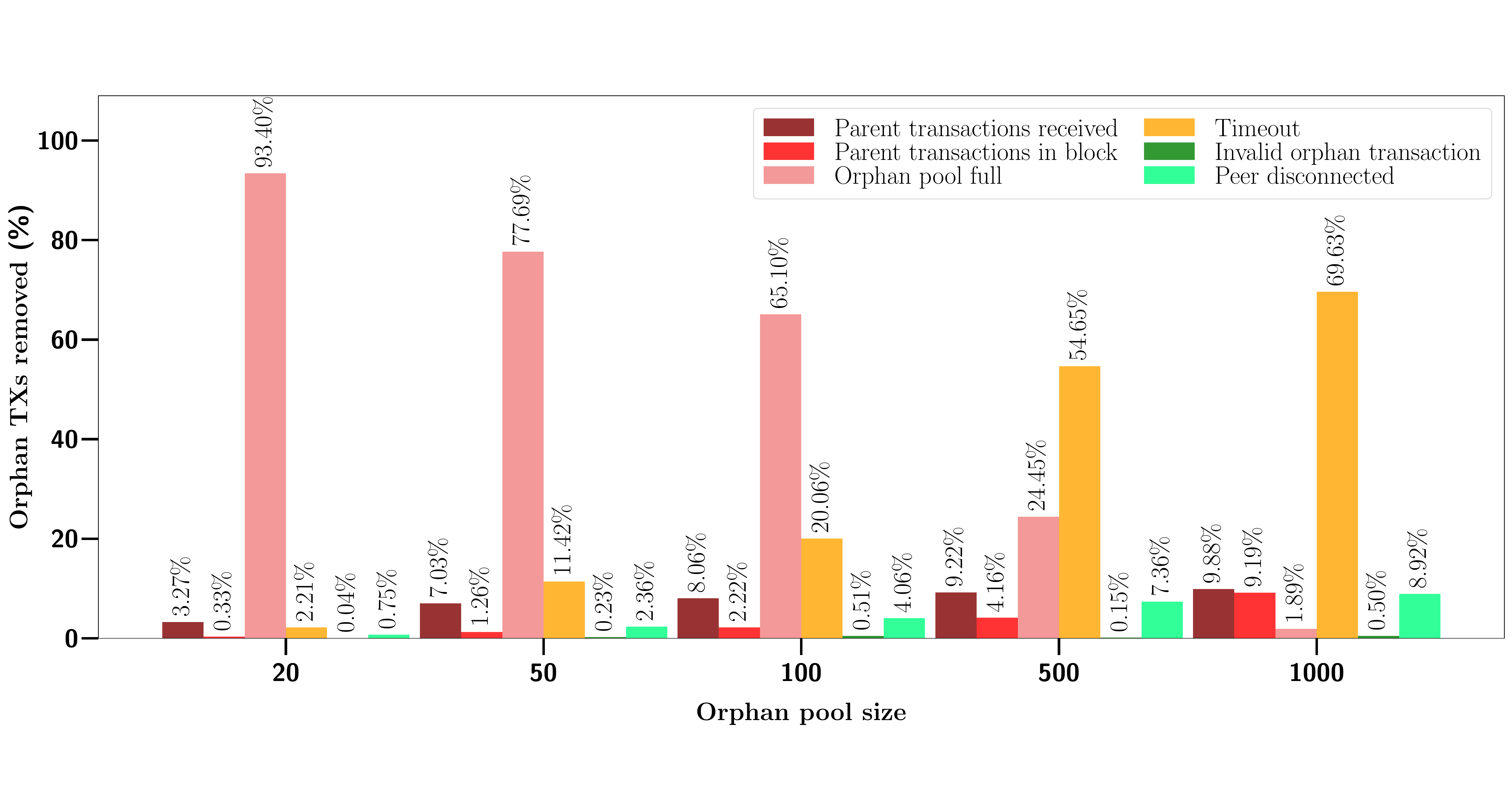}
            \caption{Fraction of orphan transactions that are removed from the orphan pool due to each of the six causes across all nodes.}
            \label{fig:tx_erase_conditions}
        \end{figure*}

    \subsection{Removal of orphan transactions from orphan pool}

        As specified in \autoref{subsec:orphantxs}, there are six different cases in which a transaction is removed from the orphan pool. In this section, we analyze the fraction of orphan transactions that are removed from the orphan pool in each case.

        Specifically, \figurename~\ref{fig:tx_erase_conditions} shows the fraction of transactions removed from the orphan transaction falling within each of the six cases across the nodes with varying orphan pool sizes.

        One trend is apparent: the major cases of transaction removal from the orphan pool are when the pool is full and when a transaction overstays its maximum allowed time in the pool. The figure clearly shows that as the size of the orphan pool increases, the major case of eviction of transactions from the orphan pool changes from the pool being full to the transactions timing out. That is, as the size of the orphan pool increases, more transactions are removed from the orphan pool due to timeout rather than a full orphan pool. In fact, one of the nodes configured with an orphan pool of size $1000$ (\ie node $N_6$) has \emph{no} transactions evicted from the orphan pool, indicating that the pool never becomes full.

        The remaining four cases contribute very little to the transaction being removed from the orphan pool. Of these, the major case that of transaction eviction from the orphan pool, across nodes, is that the node receives the missing parent it had requested from its peers. \figurename~\ref{fig:tx_erase_conditions} shows that as the size of the orphan pool increases, the fraction of orphan transactions that receive their respective missing parents gradually increases.

    \subsection{Addition of orphan transactions to orphan pool}\label{subsec:orphan_tx_add}

        \begin{figure}[t!]
            \centering
            \includegraphics[width=\columnwidth, clip]{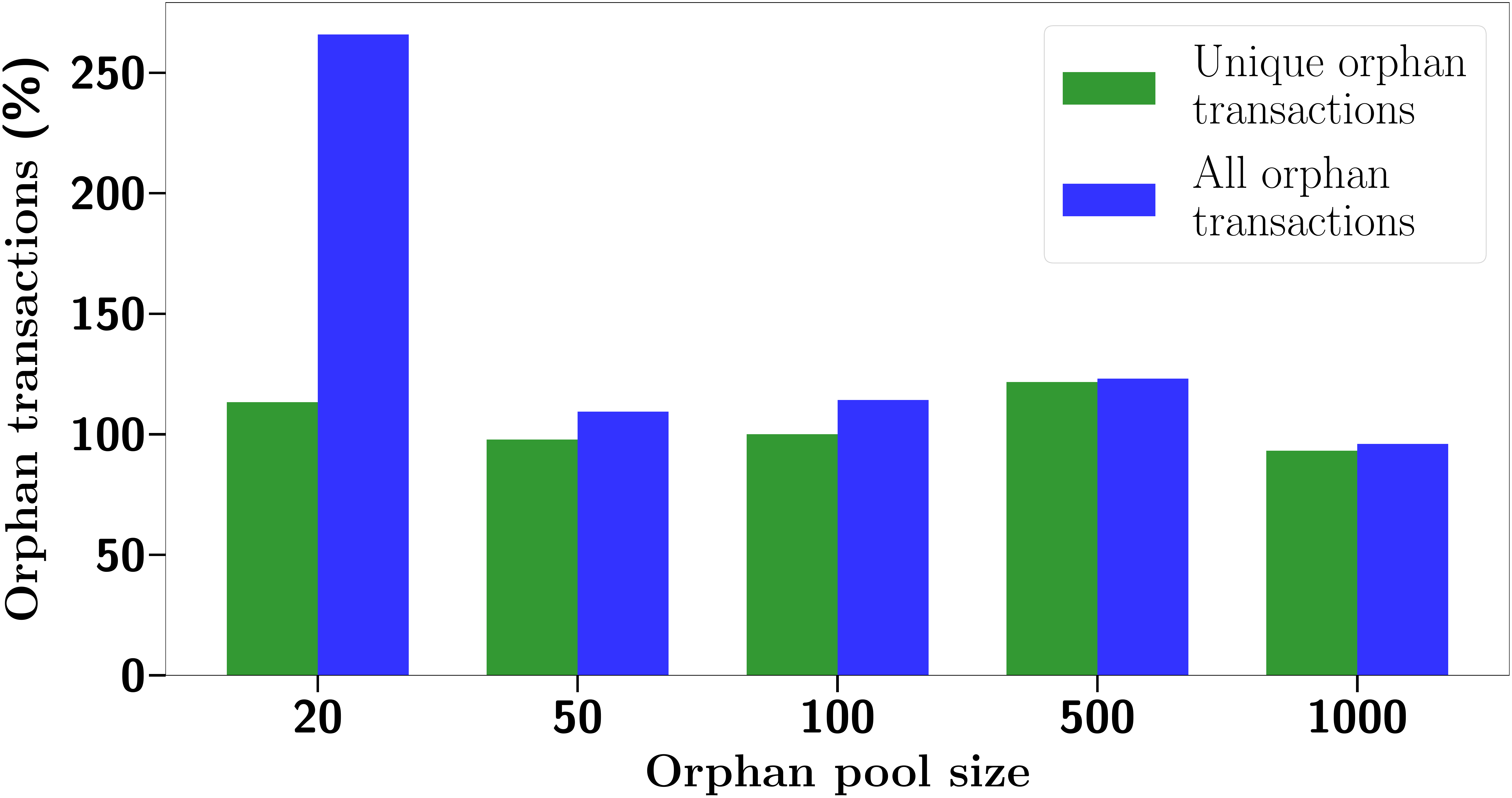}
            \caption{Number of unique and total number of orphan transactions received across nodes with varying orphan pool sizes.}
            \label{fig:num_orphan_txs}
        \end{figure}

        In the previous section, we showed that for smaller orphan pools, most transaction removals occur when the pool becomes full. However, this is not the case with orphan pools of larger sizes.  Once an orphan transaction is removed from the orphan pool without being added to the mempool (cf.~\autoref{subsec:orphantxs}), it \emph{may} be added back to the orphan pool. This happens when, after its removal from the orphan pool, a peer announces the same transaction while its parents are still missing from the mempool or the blockchain. In this section, we specifically look at the number of times a transaction may be added to the orphan pool with varying orphan pool sizes.

        To this end, the left bar in each column of \figurename~\ref{fig:num_orphan_txs} shows the \emph{unique} transactions added to the orphan pools with varying sizes. The right bar of the respective column shows the \emph{total} transactions added to the orphan pools with varying sizes.  All values are normalized to the average number of \emph{unique} transactions added to the orphan pools with a default size of $100$ over the measurement period which, on average, is $\num{57194}$ transactions.

        We observe yet another trend: for smaller orphan pool sizes, identical transactions may be added several times to the orphan pool. This is likely because smaller orphan pool fill more quickly as the number of incoming orphan transactions grows. As such, transactions need to be removed more often from the orphan pool whilst they are still orphan - a peer may re-announce a transaction that was previously removed from the orphan pool. Because the node does not have the transaction in either its mempool or the orphan pool, it accepts the transaction again to its orphan pool.
        
        When the size of the orphan pool is larger than the default size of $100$, the number of duplicate additions of transactions to the orphan pool goes down. This is likely due to the availability of space in the orphan pool for new orphan transactions; fewer transactions need to be evicted from the orphan pool. In the next section, we explain why multiple additions may pose a problem for network efficiency.

    \subsection{Network overhead}%

        \begin{figure}[t!]
            \centering
            \includegraphics[width=\columnwidth, clip]{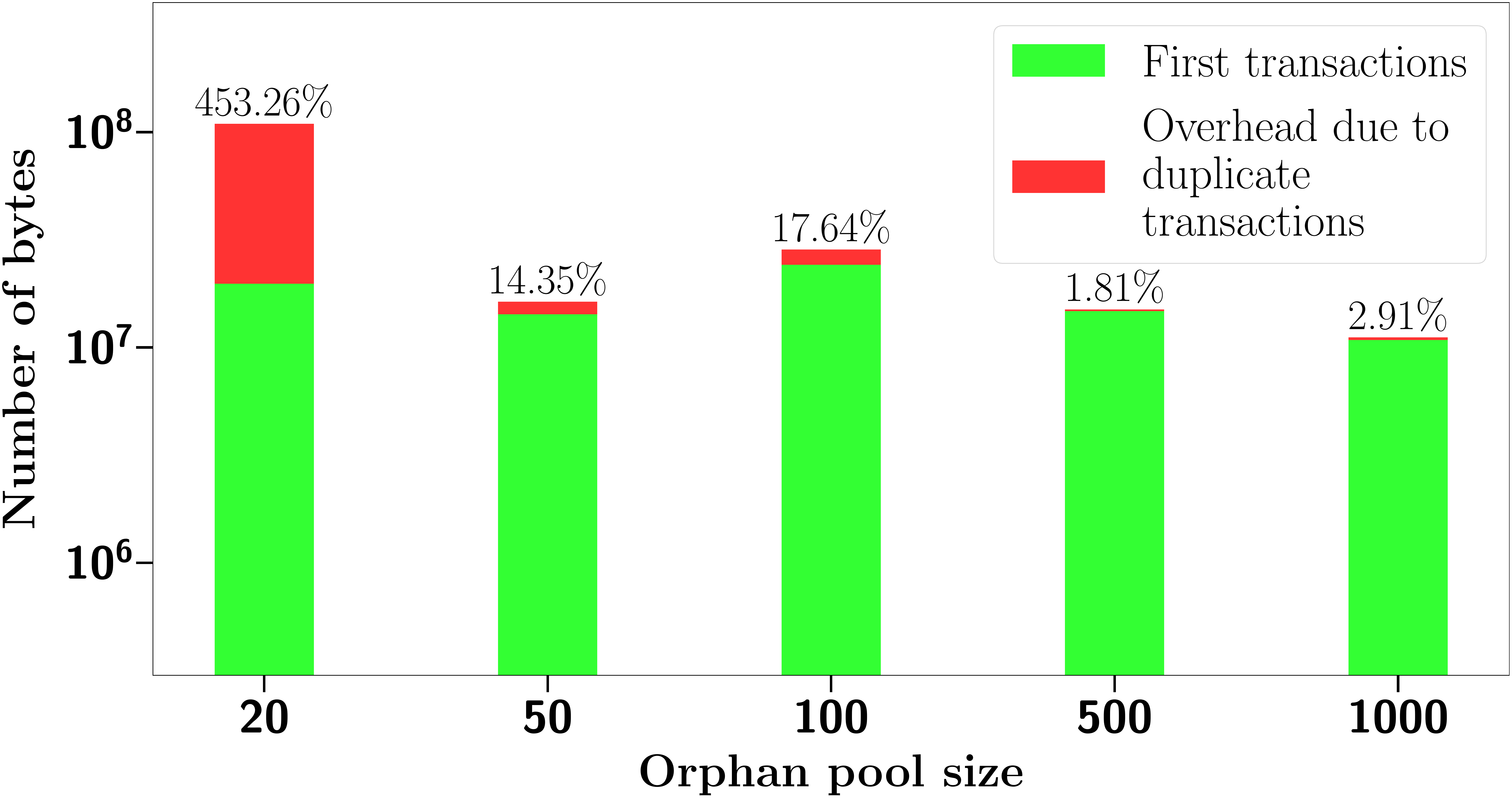}
            \caption{Network overhead incurred by nodes with varying orphan pool sizes across nodes.}
            \label{fig:network_overhead}
        \end{figure}

        We next estimate the network overhead (\ie the number of bytes received) caused by receiving duplicate orphan transactions from peers. In our experiments, each time an orphan transaction is received, we add the size of the transaction: $32$ bytes for the transaction hash in the \texttt{inv} message~\cite{inv} and $32$ bytes for the transaction hash in the \texttt{getdata} message~\cite{getdata}. Note that this provides a lower bound for the number of bytes transmitted each time a transaction is received, as the \texttt{inv} and \texttt{getdata} messages contain other fields, the total size of which would depend on the number of transactions packed in each message. We do not include this size in our calculation for simplicity. Similarly, we do not include the transport layer overhead in our estimation.

        \figurename~\ref{fig:network_overhead} shows statistics on the network overhead for duplicate orphan transactions received for the varying orphan pool sizes. The lower part of the stacked bar in each column shows the total number of bytes that are received when all \emph{unique} orphan transactions are received for the first time. The upper part of the stacked bar in the respective column shows total number of bytes received when duplicates of the orphan transactions are received; note that the $Y$-axis in this figure is logarithmic. We also provide the cost of receiving duplicate orphan transactions (above each bar) as a fraction of the cost of receiving each orphan transaction for the respective orphan pool size.

        From the figures, we see that nodes with a smaller orphan pool size incur a larger network overhead due to the repeated addition of orphan transactions to the orphan pool. On the contrary, nodes with an orphan pool of larger size incur minimal network overhead, since the number of duplicate orphan transactions received is smaller (cf. \autoref{subsec:orphan_tx_add}).

    \subsection{Performance overhead}

        \begin{table}[]
            \centering
            \small
            \resizebox{0.9\columnwidth}{!}{%
                \begin{tabular}{|c|c|c|c|c|}
                \hline
                \multirow{3}{*}{\textbf{Nodes}} & \multicolumn{2}{c|}{\textbf{Round 1}} & \multicolumn{2}{c|}{\textbf{Round 2}} \\ \cline{2-5} 
                & \textbf{\begin{tabular}[c]{@{}c@{}}Add\\ (\%)\end{tabular}} & \textbf{\begin{tabular}[c]{@{}c@{}}Remove\\ (\%)\end{tabular}} & \textbf{\begin{tabular}[c]{@{}c@{}}Add\\ (\%)\end{tabular}} & \textbf{\begin{tabular}[c]{@{}c@{}}Remove\\ (\%)\end{tabular}} \\ \hline
                $N_1$ & $18.23$ & $17.26$ & $18.71$ & $16.90$ \\ \hline
                $N_2$ & $15.26$ & $13.53$ & $15.86$ & $13.36$ \\ \hline
                $N_3$ & $18.67$ & $18.61$ & $18.37$ & $17.36$ \\ \hline
                $N_4$ & $16.37$ & $13.86$ & $15.95$ & $13.33$ \\ \hline
                $N_5$ & $18.22$ & $17.90$ & $18.67$ & $17.58$ \\ \hline
                $N_6$ & $15.85$ & $13.31$ & $16.84$ & $13.94$ \\ \hline
                \end{tabular}%
            }
            \caption{Average CPU usage of nodes with different orphan pool sizes.}
            \label{tab:cpu_usage}
        \end{table}

        Finally, we explore the CPU and memory overhead incurred by varying orphan pool sizes. We empirically measure the CPU overhead with data from Unix \texttt{procfs}, and approximate the memory overhead.  Our analysis shows that larger orphan pool sizes do not incur notable overhead for our node systems.

        \subsubsection{CPU overhead}

            The CPU overhead is observed by recording the CPU usage of the Bitcoin process every time an orphan transaction is added or removed from the orphan pool. \tablename~\ref{tab:cpu_usage} shows the \emph{average} CPU usage of the Bitcoin process over the measurement period. The table shows that the difference in the average CPU usage of the Bitcoin process is barely distinguishable among the various orphan pool sizes. We attribute this to the data structure used for the orphan pool: relevant \texttt{std::map} operations typically have worst-case logarithmic time complexity~\cite{map_emplace, map_count, map_erase}.

        \subsubsection{Memory overhead}
            The Bitcoin core maintains three data structures related to orphan transactions. The first data structure represents the orphan pool.
            Each entry for an orphan transaction in the orphan pool contains i) the hash of the transaction ($32$ bytes), ii) a pointer to the actual transaction ($16$-byte integer on $64$-bit architecture; $8$-byte integer on $32$-bit architecture; the size of this pointer is double that of an ordinary pointer because a \texttt{std::shared\_ptr} is made of $2$ pointers~\cite{shared_ptr},
            iii) the ID of the peer that sent the transaction ($8$-byte integer), iv) expiration time of the transaction ($8$-byte integer), and v) position of orphan transaction in the orphan pool ($8$-byte integer on $64$-bit architecture; $4$-byte on $32$-bit architecture).
            Considering that the transaction would be stored in the mempool anyway if it were not an orphan, each orphan transaction incurs a memory overhead of $72$ bytes on a $64$-bit architecture, and $60$ bytes on a $32$-bit architecture.

            The second data structure is used to maintain links between a missing parent and all orphan transactions that may spend from it. This efficiently resolves orphan status of all orphan transactions that depend on a missing parent once the latter is received from peers.

            Each entry in this data structure contains i) the hash of the parent ($32$ bytes), ii) the index of the parent in the orphan transaction ($4$ bytes), and iii) a pointer to the orphan transaction in the orphan pool ($8$-byte integer on $64$-bit architecture; $4$-byte integer on $32$-bit architecture). That is, each entry in this data structure takes up $36 + 8\times N$ bytes on a $64$-bit architecture, and $36 + 4\times N$ bytes on a $32$-bit architecture, where $N$ is the number of all orphan transactions that spend from a missing parent.

            It is tricky to theoretically justify a hard bound on the overhead incurred by this data structure. A transaction may spend from an arbitrary number of parents, an unknown number of which may be missing. Furthermore, not all parents may be missing at the same time, \ie a peer may not respond with \emph{all} requested missing parents at the same time. On the other hand, an arbitrary number of orphan transactions may spend from the same missing parent.

            Our empirical data, however, suggests that, orphan transactions across nodes with the varying orphan pool sizes have, on average, between 1 and 4 missing parents. where transactions across nodes with smaller pool sizes miss more parents; transactions across nodes with larger orphan pool sizes are very unlikely to miss more than $1$ parent. Indeed, more than $90\%$ of orphan transactions received by nodes configured with an orphan pool of size $1000$ miss only $1$ parent.

            Similarly, across nodes with varying orphan pool sizes, the number of missing parents that orphan transactions share is in the range $\left(0, 1\right)$ on average. For every node, more than $98\%$ of all orphan transactions received by that node share no parent.
                
            Finally, for efficient random eviction of transactions from the orphan pool when the pool is full, a list is maintained. Each entry in the list is a pointer to a transaction in the orphan pool, with an overhead of $8$-bytes for a $64$-bit architecture and $4$-bytes for a $32$-bit architecture.

        Consider, for example, a node configured with an orphan pool of size $1000$ on a $64$-bit architecture. This configuration incurs an average memory overhead of roughly $72$~KB for the first data structure, $44$~KB for the second data structure, and $8$~KB for the third data structure for an aggregated average overhead of $122$~KB, several orders of magnitude smaller than the typical memory on a modern system.

\section{Conclusion}\label{sec:conclusion}

    We have investigated circumstances under which a Bitcoin transaction is orphaned. Our data shows that orphan transactions have, on average, \emph{fewer} parents than other transactions. %
    The parents that cause transactions to become orphaned also have a lower transaction fee and %
    a larger size %
    relative to all received transactions. %
    On an individual level, the missing parents also have, on average, a lower transaction fee %
    per byte as compared to parents of all received transactions. %
    This information can be utilized by Bitcoin users to appropriately set their own transaction fees and facilitate propagation through the network.

    We have also documented the network and performance overhead incurred by orphan transactions for orphan pools of varying sizes.  Our analysis reveals that as the orphan pool size grows, more transactions are removed from the pool, not because the pool is full but because the transactions time out. This in turn reduces the duplicate addition of transactions to the orphan pool, resulting in a much smaller network overhead. Our evaluations show that the performance overhead incurred by a larger orphan pool is insignificant, and it is thus advisable to set a larger orphan pool of larger size.

    We note, in closing, that our nodes in this work are continuously connected to the Bitcoin network throughout the measurement periods.     However, prior work~\cite{imtiaz2019churn} shows that many Bitcoin nodes exhibit churn, meaning that they often lose connectivity. It would be interesting to evaluate the effect of such churn on the behavior of orphan transactions.
    
    As mentioned in {\Autoref{subsec:tx_fee, subsec:tx_size}}\, it would be interesting to determine if there may be thresholds for transaction fees and sizes, after which transactions tend to be missing, \ie they are not relayed by the network. In addition, it would be valuable to develop an algorithm, e.g., based on machine learning techniques, that can predict whether a transaction will become orphan. Finally, it would be worth investigating whether orphan transactions cause delay in propagation of blocks containing them through the Bitcoin network. 
    We leave these problems as topics for future work.
    
\section*{Acknowledgment}

    This research was supported in part by NSF under grant CCF-1563753. The authors would also like to acknowledge Sean Brandenburg for help with forward porting the log-to-file system to the newer version of the Bitcoin software.

\bibliography{bibliography}

\begin{thebibliography}{10}

\bibitem{btc_market_cap}
``Bitcoin price, charts, market cap, and other metrics.''
  \raggedright\texttt{\url{https://coinmarketcap.com/currencies/bitcoin/}}.
\newblock Online; Accessed: December 8, 2019.

\bibitem{nakamoto2008bitcoin}
S.~Nakamoto {\em et~al.}, ``Bitcoin: A peer-to-peer electronic cash system,''
  2008.

\bibitem{unconfirmed_tx}
``The complete guide to {Bitcoin} fees.''
  \raggedright\texttt{\url{https://99bitcoins.com/bitcoin/fees/}}.
\newblock Online; Accessed: December 8, 2019.

\bibitem{daily_confiremd_txs}
``Confirmed transactions per day.''
  \raggedright\texttt{\url{https://www.blockchain.com/en/charts/n-transactions?timespan=180days}}.
\newblock Online; Accessed: December 8, 2019.

\bibitem{total_txs}
``Total number of transactions.''
  \raggedright\texttt{\url{https://www.blockchain.com/charts/n-transactions-total?timespan=all}}.
\newblock Online; Accessed: December 8, 2019.

\bibitem{ron2013quantitative}
D.~Ron and A.~Shamir, ``Quantitative analysis of the full {Bitcoin} transaction
  graph,'' in {\em International Conference on Financial Cryptography and Data
  Security}, pp.~6--24, Springer, 2013.

\bibitem{ober2013structure}
M.~Ober, S.~Katzenbeisser, and K.~Hamacher, ``Structure and anonymity of the
  {Bitcoin} transaction graph,'' {\em Future internet}, vol.~5, no.~2,
  pp.~237--250, 2013.

\bibitem{fleder2015bitcoin}
M.~Fleder, M.~S. Kester, and S.~Pillai, ``Bitcoin transaction graph analysis,''
  {\em arXiv preprint arXiv:1502.01657}, 2015.

\bibitem{di2015bitconeview}
G.~Di~Battista, V.~Di~Donato, M.~Patrignani, M.~Pizzonia, V.~Roselli, and
  R.~Tamassia, ``Bitconeview: visualization of flows in the {Bitcoin}
  transaction graph,'' in {\em 2015 IEEE Symposium on Visualization for Cyber
  Security (VizSec)}, pp.~1--8, IEEE, 2015.

\bibitem{moser2013inquiry}
M.~M{\"o}ser, R.~B{\"o}hme, and D.~Breuker, ``An inquiry into money laundering
  tools in the {Bitcoin} ecosystem,'' in {\em 2013 APWG eCrime Researchers
  Summit}, pp.~1--14, Ieee, 2013.

\bibitem{greaves2015using}
A.~Greaves and B.~Au, ``Using the {Bitcoin} transaction graph to predict the
  price of {Bitcoin},'' {\em No Data}, 2015.

\bibitem{mcginn2016visualizing}
D.~McGinn, D.~Birch, D.~Akroyd, M.~Molina-Solana, Y.~Guo, and W.~J.
  Knottenbelt, ``Visualizing dynamic {Bitcoin} transaction patterns,'' {\em Big
  data}, vol.~4, no.~2, pp.~109--119, 2016.

\bibitem{androulaki2013evaluating}
E.~Androulaki, G.~O. Karame, M.~Roeschlin, T.~Scherer, and S.~Capkun,
  ``Evaluating user privacy in {Bitcoin},'' in {\em International Conference on
  Financial Cryptography and Data Security}, pp.~34--51, Springer, 2013.

\bibitem{ruffing2017valueshuffle}
T.~Ruffing and P.~Moreno-Sanchez, ``Valueshuffle: Mixing confidential
  transactions for comprehensive transaction privacy in {Bitcoin},'' in {\em
  International Conference on Financial Cryptography and Data Security},
  pp.~133--154, Springer, 2017.

\bibitem{herrera2016privacy}
J.~Herrera-Joancomart{\'\i} and C.~P{\'e}rez-Sol{\`a}, ``Privacy in {Bitcoin}
  transactions: new challenges from blockchain scalability solutions,'' in {\em
  International Conference on Modeling Decisions for Artificial Intelligence},
  pp.~26--44, Springer, 2016.

\bibitem{wang2017preserving}
Q.~Wang, B.~Qin, J.~Hu, and F.~Xiao, ``Preserving transaction privacy in
  {Bitcoin},'' {\em Future Generation Computer Systems}, 2017.

\bibitem{liu2018unlinkable}
Y.~Liu, X.~Liu, C.~Tang, J.~Wang, and L.~Zhang, ``Unlinkable coin mixing scheme
  for transaction privacy enhancement of {Bitcoin},'' {\em IEEE Access},
  vol.~6, pp.~23261--23270, 2018.

\bibitem{meiklejohn2018mobius}
S.~Meiklejohn and R.~Mercer, ``M{\"o}bius: Trustless tumbling for transaction
  privacy,'' {\em Proceedings on Privacy Enhancing Technologies}, vol.~2018,
  no.~2, pp.~105--121, 2018.

\bibitem{kawase2017transaction}
Y.~Kawase and S.~Kasahara, ``Transaction-confirmation time for {Bitcoin}: a
  queueing analytical approach to blockchain mechanism,'' in {\em International
  Conference on Queueing Theory and Network Applications}, pp.~75--88,
  Springer, 2017.

\bibitem{sompolinsky2013accelerating}
Y.~Sompolinsky and A.~Zohar, ``Accelerating {Bitcoin's} transaction processing.
  fast money grows on trees, not chains.,'' {\em IACR Cryptology ePrint
  Archive}, vol.~2013, no.~881, 2013.

\bibitem{kasahara2019effect}
S.~Kasahara and J.~Kawahara, ``Effect of {Bitcoin} fee on
  transaction-confirmation process,'' {\em Journal of Industrial \& Management
  Optimization}, vol.~15, no.~1, pp.~365--386, 2019.

\bibitem{zhu2016interactive}
Y.~Zhu, R.~Guo, G.~Gan, and W.-T. Tsai, ``Interactive incontestable signature
  for transactions confirmation in {Bitcoin} blockchain,'' in {\em 2016 IEEE
  40th Annual Computer Software and Applications Conference (COMPSAC)}, vol.~1,
  pp.~443--448, IEEE, 2016.

\bibitem{delgado2019txprobe}
S.~Delgado-Segura, S.~Bakshi, C.~P{\'e}rez-Sol{\`a}, J.~Litton, A.~Pachulski,
  A.~Miller, and B.~Bhattacharjee, ``Txprobe: Discovering {Bitcoin’s} network
  topology using orphan transactions,'' in {\em International Conference on
  Financial Cryptography and Data Security}, pp.~550--566, Springer, 2019.

\bibitem{miller2015shadow}
A.~Miller and R.~Jansen, ``Shadow-bitcoin: Scalable simulation via direct
  execution of multi-threaded applications,'' in {\em 8th Workshop on Cyber
  Security Experimentation and Test ($\{$CSET$\}$ 15)}, 2015.

\bibitem{cve_wiki}
``Cve-2012-3789.''
  \raggedright\texttt{\url{https://en.bitcoin.it/wiki/CVE-2012-3789}}.
\newblock Online; Accessed: February 12, 2020.

\bibitem{BitcoinNetProcessing}
``\texttt{netprocessing.cpp}.''
  \texttt{\url{https://github.com/bitcoin/bitcoin/blob/0.18/src/net_processing.cpp}}.
\newblock Online; Accessed: November 11, 2019.

\bibitem{bitcoinconf}
``Sample {Bitcoin} configuration file.''
  \raggedright\texttt{\url{https://github.com/MrChrisJ/fullnode/blob/master/Setup_Guides/bitcoin.conf}}.
\newblock Online; Accessed: November 11, 2019.

\bibitem{stuck_tx}
``Stuck bitcoin transaction.''
  \texttt{\url{https://bitcointalk.org/index.php?topic=5135053.0}}.
\newblock Online; Accessed: December 18, 2019.

\bibitem{bip125}
``Bip-125.''
  \raggedright\texttt{\url{https://github.com/bitcoin/bips/blob/master/bip-0125.mediawiki}}.
\newblock Online; Accessed: December 8, 2019.

\bibitem{cve_blog}
``Bitcoin orphan transactions and cve-2012-3789.''
  \raggedright\texttt{\url{https://cryptoservices.github.io/fde/2018/12/14/bitcoin-orphan-TX-CVE.html}}.
\newblock Online; Accessed: February 12, 2020.

\bibitem{dos_fix}
``Dos fix for maporphantransactions.''
  \raggedright\texttt{\url{https://github.com/bitcoin/bitcoin/pull/911}}.
\newblock Online; Accessed: February 12, 2020.

\bibitem{bitcoin_modified}
``bitcoin-releases.''
  \texttt{\url{https://github.com/nislab/bitcoin-releases/tree/icbc2020}}.
\newblock Online; Accessed: February 13, 2020.

\bibitem{imtiaz2019churn}
M.~A. Imtiaz, D.~Starobinski, A.~Trachtenberg, and N.~Younis, ``Churn in the
  {Bitcoin Network}: Characterization and impact,'' in {\em 2019 IEEE
  International Conference on Blockchain and Cryptocurrency (ICBC)},
  pp.~431--439, IEEE, 2019.

\bibitem{tx_low_fee_prop}
``A practical guide to accidental low fee transactions.''
  \raggedright\texttt{\url{https://hackernoon.com/holy-cow-i-sent-a-bitcoin-transaction-with-too-low-fees-are-my-coins-lost-forever-7a865e2e45ba}}.
\newblock Online; Accessed: December 3, 2019.

\bibitem{mempool_size}
``Is there any max limit of a mempool?.''
  \raggedright\texttt{\url{https://bitcointalk.org/index.php?topic=1714006.msg17171748\#msg17171748}}.
\newblock Online; Accessed: December 5, 2019.

\bibitem{jiang2019bitcoin}
S.~Jiang and J.~Wu, ``Bitcoin mining with transaction fees: a game on the block
  size,'' in {\em 2019 IEEE International Conference on Blockchain
  (Blockchain)}, pp.~107--115, IEEE, 2019.

\bibitem{bitcoin-logs}
``bitcoin-logs.''
  \texttt{\url{https://github.com/nislab/bitcoin-logs/tree/icbc2020}}.
\newblock Online; Accessed: February 13, 2020.

\bibitem{inv}
``Protocol documentation (inv).''
  \raggedright\texttt{\url{https://en.bitcoin.it/wiki/Protocol_documentation\#inv}}.
\newblock Online; Accessed: December 3, 2019.

\bibitem{getdata}
``Protocol documentation (getdata).''
  \raggedright\texttt{\url{https://en.bitcoin.it/wiki/Protocol_documentation\#getdata}}.
\newblock Online; Accessed: December 3, 2019.

\bibitem{map_emplace}
``\texttt{std::map::erase}.''
  \raggedright\texttt{\url{http://www.cplusplus.com/reference/map/map/emplace}}.
\newblock Online; Accessed: December 4, 2019.

\bibitem{map_count}
``\texttt{std::map::count}.''
  \raggedright\texttt{\url{http://www.cplusplus.com/reference/map/map/count/}}.
\newblock Online; Accessed: December 4, 2019.

\bibitem{map_erase}
``\texttt{std::map::erase}.''
  \raggedright\texttt{\url{http://www.cplusplus.com/reference/map/map/erase/}}.
\newblock Online; Accessed: December 4, 2019.

\bibitem{shared_ptr}
``Exploring \texttt{std::shared\_ptr}.''
  \raggedright\texttt{\url{https://shaharmike.com/cpp/shared-ptr/}}.
\newblock Online; Accessed: December 5, 2019.

\end{thebibliography}
\bibliographystyle{ieeetr}

\end{document}